\documentstyle[pra,aps,amsfonts,epsfig]{revtex}
\begin{document}
\draft
\twocolumn[\hsize\textwidth\columnwidth\hsize\csname 
@twocolumnfalse\endcsname
\title{Image and Coherence Transfer in the Stimulated Down-conversion Process} 
\author{P. H. Souto Ribeiro$^{1}$ \cite{ca}, S. P\'adua$^{2}$, 
and C. H. Monken$^{2}$}
\address{$^{1}$ Instituto de F\'{\i}sica, 
Universidade Federal do Rio de 
Janeiro, Caixa Postal 68528, Rio de Janeiro, RJ 22945-970, Brazil\\
$^{2}$ Departamento de F\'{\i}sica, 
Universidade Federal de Minas Gerais, Caixa Postal 702, 
Belo Horizonte, MG 30123-970, Brazil} 
\date{\today}
\maketitle
\begin{abstract}
The intensity transverse profile of the light produced in the process of stimulated 
down-conversion is derived. A quantum-mechanical treatment is used. 
We show that the angular spectrum of the pump laser can be transferred to the
stimulated down-converted beam, so that images can also be transferred from the
pump to the down-converted beam. We also show that the transfer can occur from the
stimulating beam to the down-converted one. Finally, we
study the process of diffraction through an arbitrarily shaped screen. For the special
case of a double-slit, the interference pattern is explicitly obtained. The visibility for the spontaneous
emitted light is in accordance with the van Cittert - Zernike theorem for incoherent light,
while the visibility for the stimulated emitted light is unity. The overall visibility is in
accordance with previous experimental results.
\end{abstract}
]
\section{INTRODUCTION}

The parametric down-conversion process has become a very important tool
on the study of the fundamentals of Quantum Mechanics. Until now, the
production of pairs of correlated photons has been the most important
feature of this process. In a cavity free configuration, down-conversion
can be accomplished in two ways: spontaneous and
stimulated down-conversion. In the first case, pairs of photons are 
spontaneously emitted, while in the stimulated
process, a second auxiliary laser is coupled to one of the down-conversion
modes. 

We will focus our attention on the stimulated down-conversion process. Even though
the great interest on the spontaneous process is due to the quantum character
of the correlations between the twin photons, it has not been demonstrated so far
any quantum property \cite{quant}of the light produced in the stimulated process.
The coherence properties of this light field has been investigated by  Mandel and 
co-workers\cite{wang}
and also in ref.\cite{eu2}. It is possible to conclude from these works
that essentially the light obtained in the stimulated down-conversion is a
superposition of light due to the spontaneous emission
process, always present, and light due to the stimulating process.
>From the point of view of the first order transverse coherence properties (field or phase coherence)
the spontaneous emitted light is incoherent in the sense discussed in ref. \cite{eu1},
and the stimulated emitted light is as coherent as the laser which induces the
emission. For the case of the longitudinal coherence properties, it was shown
that the degree of coherence of the stimulated light depends also on the pump
beam temporal coherence properties\cite{eux}. For the transverse coherence properties, however,
only a simple theory has been proposed\cite{eu2}, which does not take
into account the pump field properties.

The aim of this paper is to show how transverse propagation properties (not only
coherence properties) of the light beam produced in the stimulated down-conversion
process can be obtained from those of the pump, auxiliary laser and the parametric
interaction. The theory we present here is based on that developed in ref.\cite{monken98a}.
There, a theory was developed to take into account the transverse properties
of the correlated twin photons. We do the same, but transverse propagation effects
 (image formation and coherence transfer) are obtained in the intensity profile instead of 
in coincidence.
We obtain the intensity distribution of the stimulated down-converted beam as a function
of the pump and auxiliary laser intensity distributions. Previous experiments are
discussed and new ones are envisaged.

Finally, we would like to emphasize that the stimulated down-conversion process
is a candidate for exhibiting interesting quantum features in the recent effort
for generating multiple entangled particles. 

\section{STIMULATED INTENSITY PROFILE}

\begin{figure}
\centerline{\epsfig{file=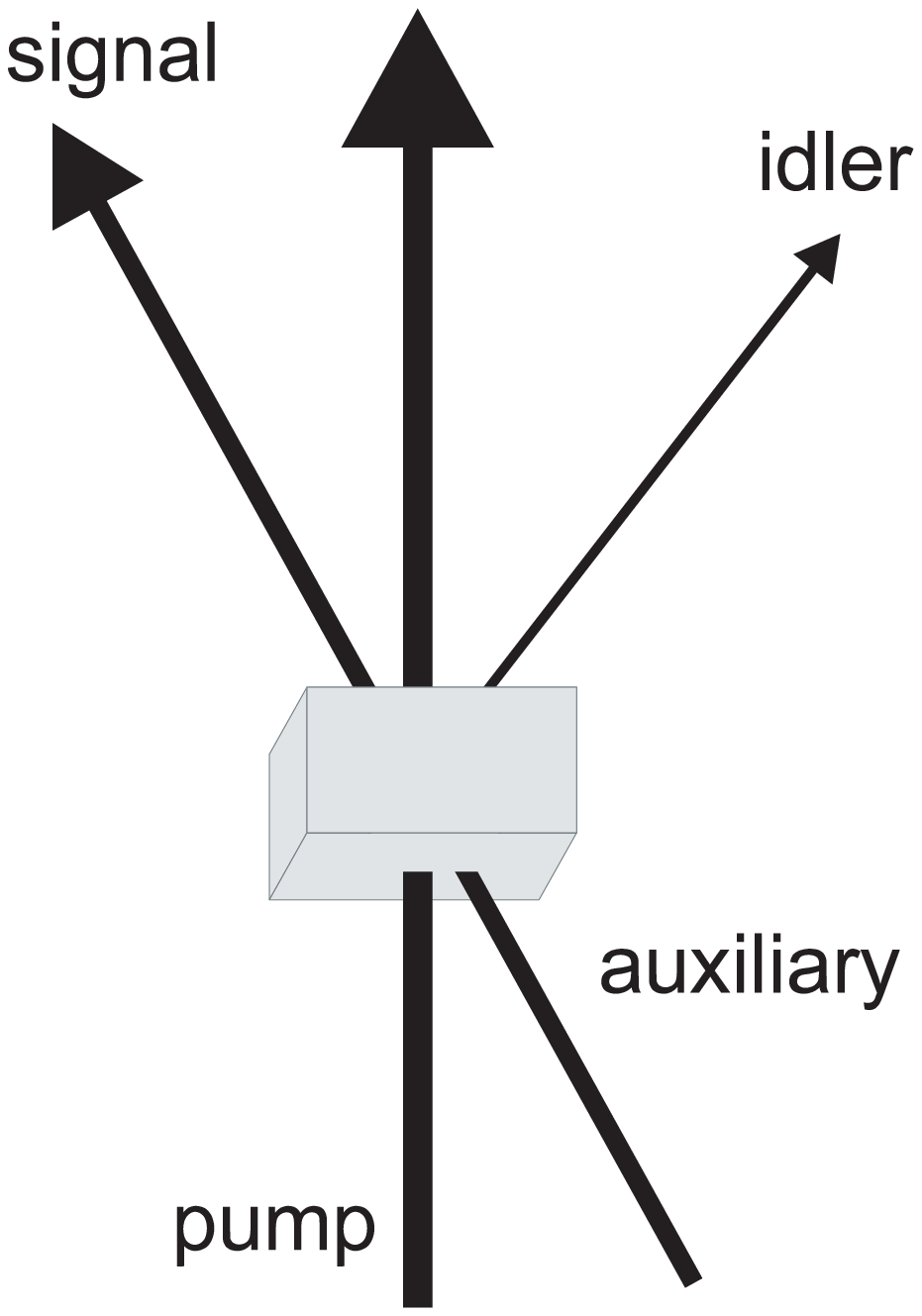, width=4cm}}
\hspace*{2cm}
\caption{Basic arrangement for the production of the stimulated
down-conversion.}
\label{fig1}
\end{figure}

A typical configuration for stimulated down-conversion is shown in
Fig. \ref{fig1}. A non-linear crystal is pumped by a laser and
produces pairs of photons in the directions labeled by signal and 
idler. A second laser is aligned along signal direction so that
its modes are coupled to signal down-conversion modes. Now,
the emission on the signal modes that are coupled to the laser
will be enhanced by stimulation. An immediate consequence of this
stimulation process is the enhancement of the idler emission.
This is a consequence of the fact that emissions are only allowed
in pairs. We are
going to proceed with the calculation of the idler beam intensity
as a function of the transverse coordinates. This will provide us
the image formed on this beam.

By using a quantum down-conversion wavefunction,
we will be able to use previous results.
For a thin nonlinear 
crystal centered at the origin and pumped along 
the $z$ direction, the state generated by SPDC  (Spontaneous Parametric
Down-conversion) in the monochromatic 
and paraxial approximations can be represented by \cite{monken98a}
\begin{eqnarray}
|\psi\rangle =&& |vac\rangle +\nonumber\\ 
&& C \int d\bbox{q}_{i} 
\int d\bbox{q}_{s} v_{p}(\bbox{q}_{i} + \bbox{q}_{s})|1;\bbox{q}_{i} 
\rangle|1;\bbox{q}_{s} \rangle,
\label{eq1}
\end{eqnarray}
where$|1;\bbox{q}\rangle$ represents a one-photon state with transverse 
wave vector component $\bbox{q}$, $v_{p}(\bbox{q})$ is the angular 
spectrum of the pump field at $z=0$, and C is a constant.

For the stimulated parametric down-conversion an analogous wavefunction
is given by\cite{wang} :

\begin{eqnarray}
|\psi\rangle =&& |v_{s}(\bbox{q})\rangle |vac\rangle +\nonumber\\ 
&& C \int d\bbox{q}_{i} 
\int d\bbox{q}_{s} v_{p}(\bbox{q}_{i} + \bbox{q}_{s})|1;\bbox{q}_{i} 
\rangle a^{\dagger}(\bbox{q}_{s})|v_{s}(\bbox{q})\rangle,
\label{eq2}
\end{eqnarray}
where$|v_{s}(\bbox{q})\rangle$ is a multimode coherent state in the continuous
mode representation \cite{mwolf} with transverse 
wave vector component $\bbox{q}$, and $v_{s}(\bbox{q})$ is the angular 
spectrum of the stimulating field at $z=0$. 

The idler beam intensity is given by the second order correlation
function :

\begin{eqnarray}
I(\bbox{r}_{i})=&&\langle 
E^{(-)}_{i}(\bbox{\rho}_{i},z)E^{(+)}_{i}(\bbox{\rho}_{i},z)\rangle,
\label{eq3}
\end{eqnarray}
where $\bbox{\rho}_{i}$ is the transverse component of $\bbox{r}_{i}$
and $z$ is the longitudinal one. $E^{(+)}_{i}(\bbox{\rho}_{i},z)$ is
the electric field operator given by \cite{monken98a}:

\begin{eqnarray}
E^{(+)}_{i}(\bbox{\rho}_{i},z)= \int d\bbox{q}_{i} a(\bbox{q}_{i})
\exp \left[i\left(\bbox{q}_{i}\cdot\bbox{\rho}_{i}-
\frac{q_{i}^{2}}{2k_{i}}z\right)\right].
\label{eq4}
\end{eqnarray}

The intensity distribution is then given by :

\begin{eqnarray}
\label{eq5}
I(\bbox{r}_{i}) 
= && |C|^{2}\int d\bbox{q}_{i} \int d\bbox{q}_{s} 
\int d\bbox{q'}_{i} \int d\bbox{q'}_{s} 
v_{p}(\bbox{q}_{i} + \bbox{q}_{s}) \times \\ \nonumber
&& v^{\star}_{p}(\bbox{q'}_{i} + \bbox{q'}_{s})
\exp \left[i\left(\bbox{q}_{i}\cdot\bbox{\rho}_{i}-
\frac{q_{i}^{2}}{2k_{i}}z\right)\right] \times \\ \nonumber
&&\exp \left[-i\left(\bbox{q'}_{i}\cdot\bbox{\rho}_{i}-
\frac{q_{i}^{'2}}{2k_{i}}z\right)\right]\times \\ \nonumber
&&
\langle v_{s}(\bbox{q})| a(\bbox{q'}_{s})
a^{\dagger}(\bbox{q}_{s}) |v_{s}(\bbox{q})\rangle .
\end{eqnarray}

By means of the commutation relation

\begin{equation}
\label{eq6}
[a(\bbox{q'}_{s}),a^{\dagger}(\bbox{q}_{s})] = 
\delta(\bbox{q'}_{s} - \bbox{q}_{s}),
\end{equation}
and the Fourier integral representation of the pump beam angular
spectrum :

\begin{equation}
\label{eq8}
v_{p}(\bbox{q}_{i} + \bbox{q}_{s})\ = c \int d\bbox{\rho}
{\cal W}_{p}(\bbox{\rho}) \\ \nonumber
\exp [-i(\bbox{q}_{i} + \bbox{q}_{s})\cdot \bbox{\rho}],
\end{equation}
eq.\ref{eq5} can be written as

\begin{eqnarray}
\label{eq9}
I(\bbox{r}_{i})
= && |C|^{2}\int d\bbox{q}_{i} \int d\bbox{q}_{s} 
\int d\bbox{q'}_{i} \int d\bbox{q'}_{s} \times \\ \nonumber
&& \int d\bbox{\rho}
{\cal W}_{p}(\bbox{\rho})
\exp [-i(\bbox{q}_{i} + \bbox{q}_{s})\cdot \bbox{\rho}] \times 
\\ \nonumber
&& \int d\bbox{\rho}'
{\cal W}_{p}^{\star}(\bbox{\rho}') 
\exp [i(\bbox{q'}_{i} + \bbox{q'}_{s})\cdot \bbox{\rho}']
\times \\ \nonumber
&& \exp \left[i\left(\bbox{q}_{i}\cdot\bbox{\rho}_{i}-
\frac{q_{i}^{2}}{2k_{i}}z\right)\right] \times \\ \nonumber
&& \exp \left[-i\left(\bbox{q'}_{i}\cdot\bbox{\rho}_{i}-
\frac{q_{i}^{'2}}{2k_{i}}z\right)\right]\times \\ \nonumber
&& 
[\delta(\bbox{q'}_{s} - \bbox{q}_{s})  + 
v^{\star}_{s}(\bbox{q}_{s})v_{s}(\bbox{q'}_{s})].
\end{eqnarray}

Now, let us perform the integrals in momentum variables :

\begin{mathletters}
\label{eq10}
\begin{eqnarray}
&&\int d\bbox{q}_{i}\exp \left\{i\left[\bbox{q}_{i}\cdot
(\bbox{\rho}_{i}-\bbox{\rho})-
\frac{q_{i}^{2}}{2k_{i}}z\right]\right\} = \\ \nonumber
&& = \exp \left[i|\bbox{\rho}_{i}-\bbox{\rho}|^{2}
\frac{k_{i}}{2z}\right]
\end{eqnarray}
\label{eq10b}
\begin{equation}
\int d\bbox{q}_{s}  \exp(-i\bbox{q}_{s}\cdot \bbox{\rho}) 
v_{s}^{\star}(\bbox{q}_{s}) = {\cal W}_{s}^{\star}(\bbox{\rho}).
\end{equation}
\end{mathletters}

Using eqs. \ref{eq10} in eq. \ref{eq9}, we obtain 

\begin{eqnarray}
\label{eq12}
I(\bbox{r}_{i}) = &&  |C|^{2} \biggr\{ \int d\bbox{\rho} 
\left|{\cal W}_{p}(\bbox{\rho})\right|^{2} +
\\ \nonumber
&& \biggr| \int d\bbox{\rho}
{\cal W}_{p}(\bbox{\rho}) {\cal W}_{s}^{\star}(\bbox{\rho})
\exp \left[i |\bbox{\rho}_{i} - \bbox{\rho}|^{2}\frac{k_{i}}{2z}\right]
\biggr|^{2} \biggr\}.
\end{eqnarray}

First term in eq.\ref{eq12} is constant relatively to detection position
$\bbox{\rho}_{i}$. This is the spontaneous emission term. Even without performing
the integral in the second term, we can interpret it as being a consequence
of the stimulation as it depends on the stimulating laser properties through
${\cal W}_{s}^{\star}(\bbox{\rho})$.
This result stresses a known feature of the stimulated
process. The light beam is a superposition of the light produced in the
spontaneous  (first term) plus the light produced in the stimulated emission 
process  (second term).

The stimulated emission term, shows that the intensity profile is given by
the product of the pump and the auxiliary lasers transverse intensity 
distributions at the crystal ($z = 0$), propagated by a Fresnel propagator
to the idler detection plane. 
It works as if the stimulating field were diffracted by the transverse profile
of the pump field and vice-versa. This is equivalent to a photon-photon scattering
process.
This term can be much stronger than
the first one, depending on the auxiliary laser intensity, and then dominating
the idler intensity profile.

For example, if the auxiliary laser has a constant transverse intensity distribution,
an image formed by the the pump beam is transferred to the idler. Fig.\ref{fig3}
illustrates this situation when we use a mask.
This is analogous
to the image transfer to the fourth order correlation function demonstrated
in ref.\cite{monken98a}, but now the image is transferred to the second
order correlation function (intensity).

If the pump beam profile is constant,  one image formed by the auxiliary laser
beam is transferred to the idler. However, in this case it is the image corresponding
to the complex conjugate of the auxiliary laser field. We have
phase conjugation, which can be used in many ways for treating images and
laser beams.

\section{DOUBLE-SLIT EXPERIMENT WITH STIMULATED DOWN-CONVERSION}

The degree of transverse coherence of the light produced in the
stimulated down-conversion process has been studied experimentally
in ref.\cite{eu2}. By performing double-slit experiments, it was
shown that this light source behaves as a superposition of
two kinds of light sources. One, incoherent, in the sense that the
spatial correlation in the source is of the kind $\delta (\bbox{r})$,
if $\bbox{r}$ is the coordinate of one point in the source\cite{eu1}. Another,
coherent, so that the spatial correlation over the source surface is
always unity. However, only a simple explanation was given.

Now, we are able to use the theory introduced in ref.
\cite{monken98a} to give a more general explanation for that
experiment. The advantage of the present treatment is that the
calculation can be carried out for an arbitrary shape of the
screen and not only limited to the case of the double-slits.
Besides, it takes into account the spatial properties of the pump
and stimulating beams in a more general way.

\begin{figure}
\centerline{\epsfig{file=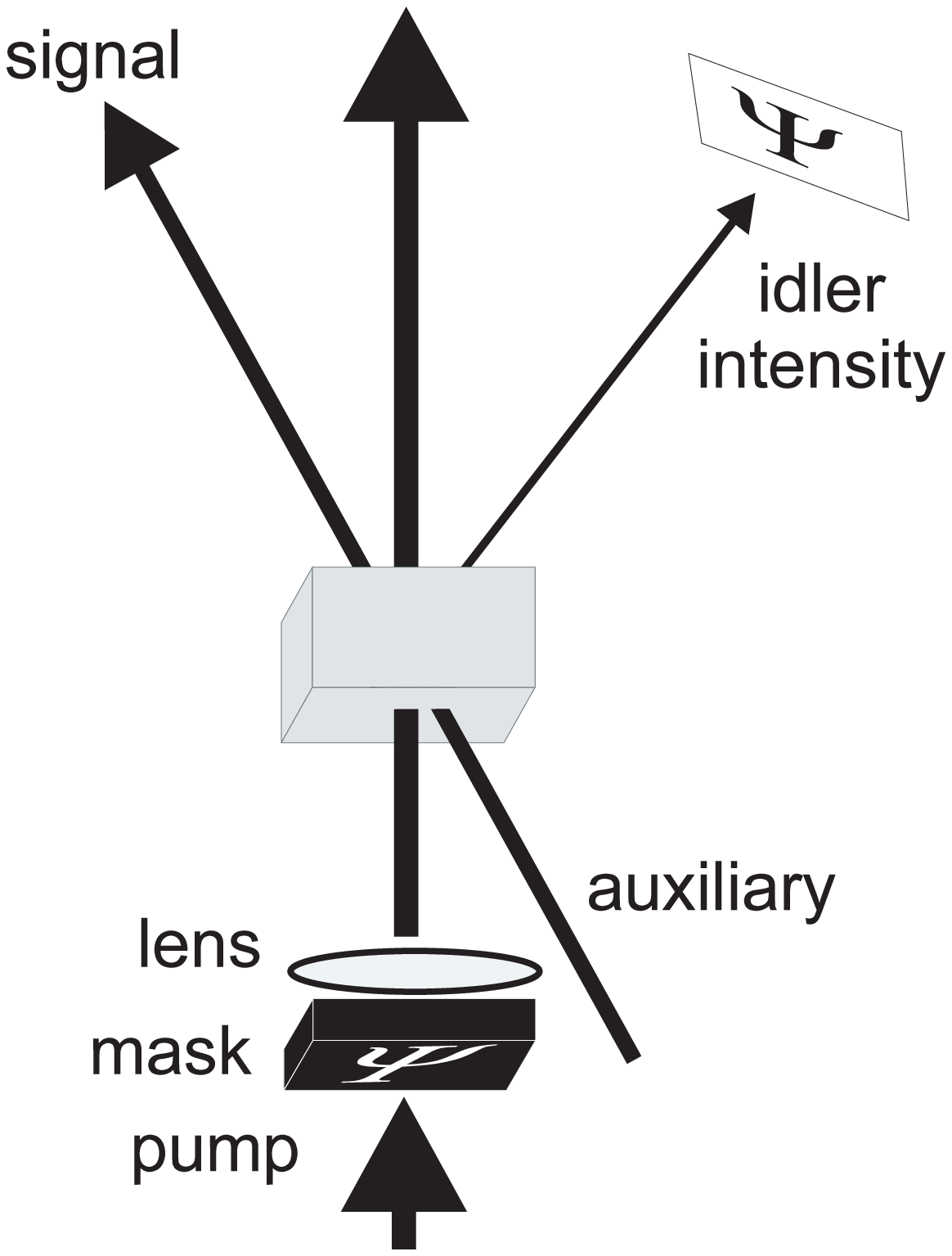, width=4cm}}
\hspace*{2cm}
\caption{Image transfer from the pump to the down-conversion beam.}
\label{fig2}
\end{figure}

\begin{figure}
\centerline{\epsfig{file=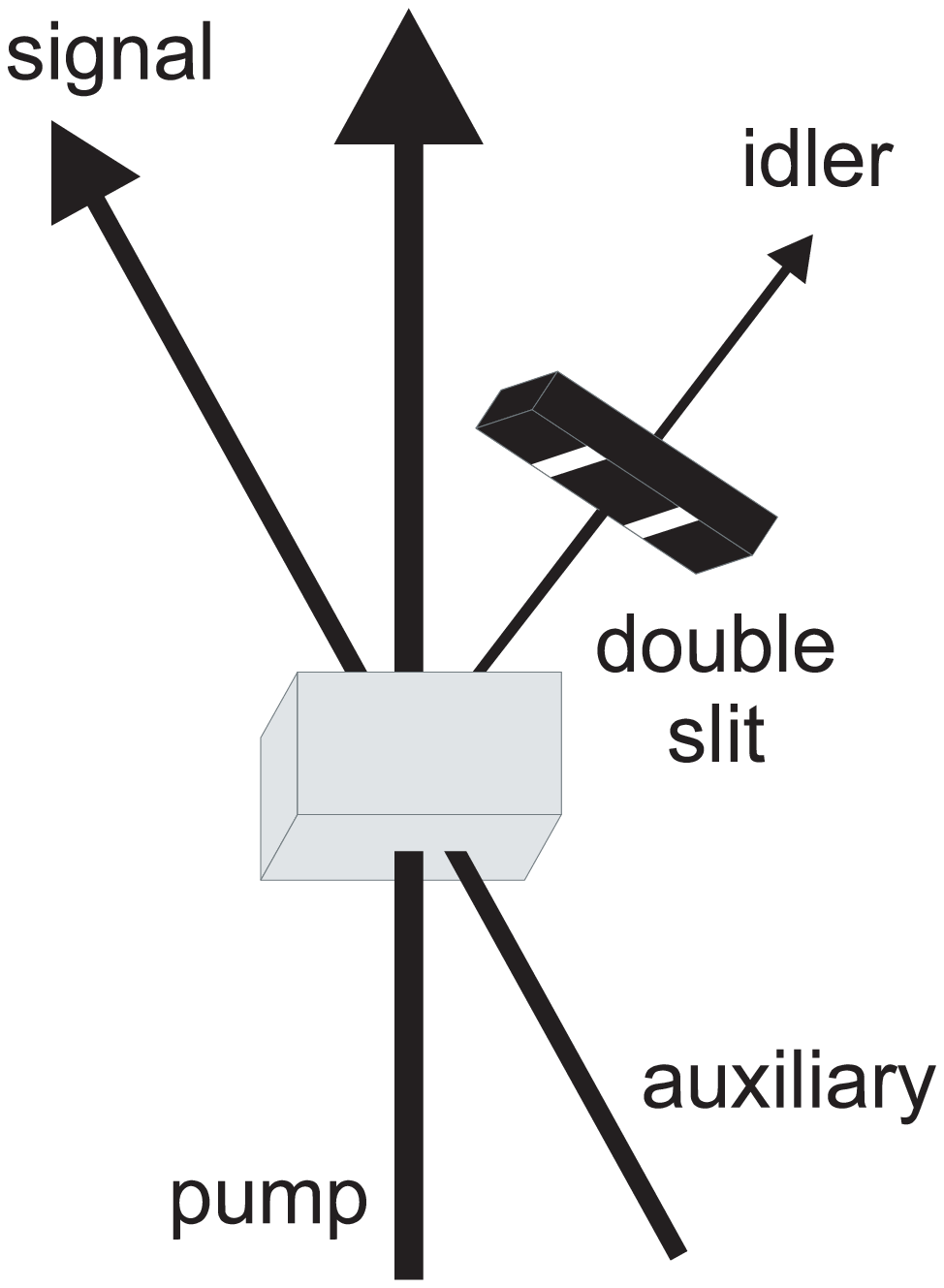, width=4cm}}
\hspace*{2cm}
\caption{Scattering of the stimulated down-conversion by a double-slit.}
\label{fig3}
\end{figure}

Let us think about the situation described in Fig. \ref{fig2}.
The stimulating laser is aligned with the signal beam  and the idler
is passed through an aperture on the screen ``S'' whose shape is 
given by the function ${\cal A}(\rho)$ defined in the plane of the 
screen. One special case of the aperture is the double-slit shape.
We want to calculate the intensity distribution of the idler beam
after the screen. We will suppose that the distances crystal-screen 
and screen-detection plane are large enough to
use the paraxial approximation. The intensity is given in the same
way as before, by 

\begin{eqnarray}
\label{eq13}
I(\bbox{r}_{i})=&&\langle 
E^{(-)}_{i}(\bbox{\rho}_{i},z)E^{(+)}_{i}(\bbox{\rho}_{i},z)\rangle,
\end{eqnarray}
as well as the state of the down-converted light :

\begin{eqnarray}
\label{eq14}
|\psi\rangle =&& |v_{s}(\bbox{q})\rangle |vac\rangle +\nonumber\\ 
&& C \int d\bbox{q}_{i} 
\int d\bbox{q}_{s} v_{p}(\bbox{q}_{i} + \bbox{q}_{s})|1;\bbox{q}_{i} 
\rangle a^{\dagger}(\bbox{q}_{s})|v_{s}(\bbox{q})\rangle.
\end{eqnarray}

The electric field operator, however, is changed in order to
take into account the propagation through the aperture. This is
done in the same way as in ref.\cite{ed1} :

\begin{eqnarray}
\label{eq15}
E^{(+)}_{i}(\bbox{\rho}_{i},z)= && \exp(ik_{i}z)
\int d\bbox{q}_{i} \int d\bbox{q'} a(\bbox{q'}) T(\bbox{q}_{i} - \bbox{q'})
\times \\ \nonumber
&& \exp \left\{ i \left[ \bbox{q}_{i}\cdot\bbox{\rho}_{i}-
\frac{q_{i}^{2}}{2k_{i}}(z - z_{A}) -
\frac{q^{'2}}{2k_{i}}z_{A} \right] \right\},
\end{eqnarray}
where $T(\bbox{q})$ is the Fourier transform of ${\cal A}(\bbox{\rho})$,
$z_{A}$ is the longitudinal coordinate of the screen and $z$ is the
longitudinal coordinate of the detection plane. The crystal is located at
z=0.

The intensity after the screen is then given by 

\begin{eqnarray}
\label{eq16}
I(\bbox{r}_{i}) 
= &&\biggr|C\int d\bbox{q}_{i} \int d\bbox{q'} \int d\bbox{q}_{s} 
v_{p}(\bbox{q}_{i} + \bbox{q}_{s})T(\bbox{q'} - \bbox{q}_{i}) 
\times \\ \nonumber
&& \exp \left[i\left(\bbox{q'}\cdot\bbox{\rho}_{i}-
\frac{q^{'2}}{2k_{i}}(z - z_{A})- \frac{q_{i}^{2}}{2k_{i}}z_{A}
\right)\right]\times \\ \nonumber
&& a^{\dagger}(\bbox{q}_{s})|v_{s}(\bbox{q})\rangle 
|0;\bbox{q}_{i}\rangle \biggr|^{2}.
\end{eqnarray}

In order to perform the calculation, it is necessary to follow
the same procedure as in the previous section, using the 
commutation relation

\begin{equation}
\label{eq17}
[a(\bbox{q'}_{s}),a^{\dagger}(\bbox{q}_{s})] = 
\delta(\bbox{q'}_{s} - \bbox{q}_{s}),
\end{equation}

and performing integrals in momentum degrees of freedom

\begin{mathletters}
\begin{eqnarray}
\label{eq19}
&& \int d\bbox{q}_{s}  \exp(-i\bbox{q}_{s}\cdot \bbox{\xi}) 
v_{s}^{\star}(\bbox{q}_{s}) = {\cal W}_{s}^{\star}(\bbox{\xi}) , \\
\label{eq19a}
&& \int d\bbox{q}_{i} \exp[i \bbox{q}_{i}\cdot(\bbox{\eta} - \bbox{\xi})]
\exp \left[-i \frac{q_{i}^{2}}{2k_{i}}z_{A}\right] = \\ \nonumber
&& = \exp\left[i\left(|\eta - \xi|^{2}
\frac{k_{i}}{2z_{A}}\right)\right] ,\\ \nonumber
\label{eq19b}
&& \int d\bbox{q} \exp[i \bbox{q}\cdot(\bbox{\rho}_{i} - \bbox{\eta})]
\exp \left[-i \frac{q^{2}}{2k_{i}}(z - z_{A})\right] = \\
&& = \exp\left[i|\rho_{i} - \eta|^{2}
\frac{k_{i}}{2(z - z_{A})}\right].
\end{eqnarray}
\end{mathletters}

The intensity is now 

\begin{eqnarray}
\label{eq20}
&& I(\bbox{r}_{i}) = 
|C|^{2}\biggr\{  \int d\bbox{\xi} \left| {\cal W}_{p}(\bbox{\xi})\right|^{2}
\times \\ \nonumber
&& \biggr| \int d\bbox{\eta} {\cal A}(\bbox{\eta})
\exp\left[i|\eta - \xi|^{2}
\frac{k_{i}}{2z_{A}}\right] 
\times \\ \nonumber
&& \exp\left[i|\rho_{i} - \eta|^{2}
\frac{k_{i}}{2(z - z_{A})}\right]\biggr|^{2} +
\\ \nonumber
&& \biggr| \int d\bbox{\xi}\int d\bbox{\eta}
{\cal W}_{p}(\bbox{\xi}) {\cal W}_{s}^{\star}(\bbox{\xi}) {\cal A}(\bbox{\eta})
\times \\ \nonumber
&& \exp\left[i|\eta - \xi|^{2}
\frac{k_{i}}{2z_{A}}\right] 
\exp\left[i|\rho_{i} - \eta|^{2}
\frac{k_{i}}{2(z - z_{A})}\right]
\biggr|^{2}\biggr\}.
\end{eqnarray}

We cannot proceed with the calculation without specifying the spectral and aperture
functions. However, if we take the Fraunhofer limit, the phase functions become :

\begin{eqnarray}
\label{eq201}
&& \exp\left(-i|\bbox{\eta} - \bbox{\xi}|^{2}\frac{k_{i}}{2z_{A}}\right)
\rightarrow 
\exp\left(-i \bbox{\eta} \cdot \bbox{\xi} \frac{k_{i}}{z_{A}}\right);
\\ \nonumber
&& \exp\left[-i|\bbox{\rho} - \bbox{\eta}|^{2}\frac{k_{i}}{2(z-z_{A})}\right]
\rightarrow 
\exp\left(-i \bbox{\rho} \cdot \bbox{\eta} \frac{k_{i}}{z-z_{A}}\right),
\end{eqnarray}
and the intensity is changed to

\begin{eqnarray}
\label{eq202}
I(\bbox{r}_{i})
=  && |C|^{2}\biggr\{  \int d\bbox{\xi} \left| {\cal W}_{p}(\bbox{\xi})\right|^{2}
|T(\beta_{1}\bbox{\xi} + \beta_{2}\bbox{\rho}_{i})|^{2} + \\ \nonumber
&& \biggr| \int d\bbox{\xi}  {\cal W}_{p}(\bbox{\xi}) {\cal W}_{s}^{\star}(\bbox{\xi})
T(\beta_{1}\bbox{\xi} + \beta_{2}\bbox{\rho}_{i})\biggr|^{2} \biggr\},
\end{eqnarray}

where 

\begin{eqnarray}
\label{eq202a}
\beta_{1} = \frac{k_{i}}{2z_{A}} \,\,\,\,\,\mbox{and} \\ \nonumber
\beta_{2} = \frac{k_{i}}{2(z - z_{A})}.
\end{eqnarray}

This result shows that the intensity pattern after a screen with
arbitrary shape, in the Fraunhofer diffraction limit  (which is
quite acceptable for experimental situations) is given by the
superposition of two terms. The first one is a convolution
between the pump beam intensity distribution and the square modulus
of the Fourier transform of the aperture function. This corresponds
to an incoherent scattering through the aperture. The second term is
the square modulus of a convolution between two field amplitudes and
the Fourier transform of the aperture function. One of the fields is
the pump and another is the conjugate of the auxiliary laser.

Thus, no matter what is the shape of the aperture, the scattered light
will be composed of an incoherent plus a coherent part. Let us now
carry out the calculation of the idler intensity distribution for a particular
shape of the aperture.

\section{1D uniform source and thin double-slits}

To illustrate the power of  the above result, let us calculate explicitly the
intensity profile for a one-dimensional case with uniform pump and
auxiliary lasers intensity distributions. Also, let us consider thin
double-slits separated by a distance $2d$, so that we have the following configuration :

\begin{eqnarray}
\label{eq203}
&&{\cal W}_{p}(\bbox{x}) = 0; if -a > x > a ; \\ \nonumber 
&&{\cal W}_{p}(\bbox{x}) =  w_{p}; if -a < x < a ; \\ \nonumber 
&&{\cal W}_{s}(\bbox{x}) = 0; if -a > x > a ; \\ \nonumber 
&&{\cal W}_{s}(\bbox{x}) =  w_{s}; if -a > x > a ; \\ \nonumber 
&&{\cal A}(\bbox{x}) = \delta(x+d) + \delta(x-d).
\end{eqnarray}

Then the intensity after the slits is given by :

\begin{eqnarray}
\label{eq204}
I(x_{i}) =  && |C|^{2}w_{p}^{2}4a\biggr\{ \left[1 + \frac{\sin(2\beta_{1}da)}{2\beta_{1}da} \cos(2\beta_{2}dx_{i})\right]
+ \\ \nonumber
&& w_{s}^{2}2a\left[\frac{\sin(\beta_{1}da)}{\beta_{1}da}\right]^{2} [1 + \cos(2\beta_{2}dx_{i})]\biggr\}.
\end{eqnarray}

By using the following definitions : 
\begin{eqnarray}
\label{eq206}
I_{0} = I_{SP} + I_{ST}; \\ \nonumber
\mu = \frac{I_{SP}\mu_{SP} + I_{ST}\mu_{ST}}{I_{0}}; \\ \nonumber
I_{SP} = 4a |C|^{2} w_{p}^{2}; \\ \nonumber
I_{ST} = 8a^{2}\left[\frac{\sin(\beta_{1}da)}{\beta_{1}da}\right]^{2} |C|^{2} w_{p}^{2} w_{s}^{2}; \\ \nonumber
\mu_{SP} = \frac{\sin(2\beta_{1}da)}{2\beta_{1}da}; \\ \nonumber
\mu_{ST} = 1,
\end{eqnarray}
the intensity distribution can be put in the form :
\begin{equation}
\label{eq205}
I(x_{i}) = I_{0}[1 + \mu\cos(2\beta_{2}dx_{i})].
\end{equation}

First of all, note that the overall visibility for the interference fringes $\mu$ is
in complete agreement with that of ref. \cite{eu2} which has been experimentally
tested. Note also that the visibility of the spontaneous emission part $\mu_{SP}$
is exactly the visibility given by the van Cittert-Zernike theorem for spatially
incoherent light, which were also checked for the down-conversion in ref.\cite{eu1}.

The stimulated intensity $I_{ST}$ is a diffraction pattern of a coherent field
through a slit with the source dimensions $2a$. This stresses the photon-photon
character of the interaction between pump and auxiliary lasers.

\section{CONCLUSION}

As a conclusion, we have shown how the transverse properties  (angular spectrum)
of the light produced in the stimulated down-conversion process can be obtained
from those of the pump, auxiliary laser beam, parametric interaction and the
shape of apertures that can be eventually placed in any one of these beams. 
It is shown that images in the pump field, produced by a mask for example, 
can be transferred to the down-converted field. Images in the stimulating
field can also be transferred to the down-converted field, but through the
conjugated field.

Some curious possibilities can be studied experimentally in further
works. The first is to produce interference fringes in the down-converted
field, placing slits in the pump beam. The second is to produce
interference fringes in the down-converted field placing slits in
the stimulating laser. Another one, is to produce interference fringes in the
idler beam, by placing one part of a double-slit in the pump and 
another part in the auxiliary laser, as in ref.\cite{ed1} 

Our results were also used to discuss a double-slit experiment with stimulated
down-conversion light, showing generally, the influence of the pump and auxiliary lasers angular properties
on the interference effects. We show that the theoretical description
explains previous experimental results\cite{eu2}, where interference fringes
were a consequence of two contributions, one due to the
spontaneous emission and another due to the stimulated emission
process.

We believe this problem is already interesting from the fundamental point of view.
However some applications can be envisaged. For example, it is possible
to produce a light signal containing one image superposed to a background. As the image
information is contained in the coherent part of the field and the background is
produced by the incoherent field, it is possible to filter image information from
the background noise with spatial filters. Another interesting point is that the
image transfer can occur with beams of different wavelengths. Many 
experimental situations can be thought, by controlling pump and auxiliary
laser angular properties and intensities. Finally, quantum properties of
this kind of field can also be studied with the same formalism.

\section{ACKNOWLEDGMENT}
The authors acknowledge financial support from the Brazilian 
agencies CNPq, FINEP, PRONEX, FAPERJ and FAPEMIG. 

\end{document}